\documentclass[a4paper,english,11pt]{article}
\usepackage{graphicx}
\usepackage{amssymb}
 \usepackage[dvips]{epsfig} 
 \usepackage{a4,amssymb,epsfig,array,cite}

\usepackage{amsfonts,bm,amssymb,euscript,array,babel}

\renewcommand{\title}[1]{\vspace{10mm}\noindent{\Large{\bf #1}}\vspace{8mm}}
\newcommand{\authors}[1]{\noindent{\large #1}\vspace{5mm}}
\newcommand{\address}[1]{{\itshape #1\vspace{2mm}}}

\usepackage{graphicx}

\newcommand{\be}{\begin{equation}}
\newcommand{\ee}{\end{equation}}
\newcommand{\eq}[1]{(\ref{#1})}
\def\nn{\nonumber}
\def\bea{\begin{eqnarray}}
\def\eea{\end{eqnarray}}
\def\cM{{\cal M}}
\def\cC{{\cal C}}
\def\cA{{\cal A}}
\def\cH{{\cal H}}

\def\R{{\mathbb R}}

\def\Tr{{\rm Tr}}
\def\tr{{\rm tr}}

\def\a{\alpha}
\def\d{\delta}
\def\g{\gamma}

\def\one{\mbox{1 \kern-.59em {\rm l}}}

\begin{document}

\sloppy

\begin{center}

\title{Matrix Models, Emergent Gravity, and Gauge Theory}

\authors{Harold Steinacker}

\address{Fakult\"at f\"ur Physik, Universit\"at Wien\\
 Boltzmanngasse 5, A-1090 Wien, Austria\\
E-mail:~\textsf{harold.steinacker@univie.ac.at}}


\begin{abstract}

Matrix models of Yang-Mills type induce an effective gravity
theory on 4-dimensional branes, which are considered as
models for dynamical space-time. We review
recent progress in the understanding of this emergent gravity.
The metric is not fundamental but
arises effectively in the semi-classical limit, along with
nonabelian gauge fields. This leads to 
a mechanism for protecting certain geometries from corrections
due to the vacuum energy.

\end{abstract}

\end{center}

\section{Background and motivation}

Quantum field theory and general relativity (GR) provide
the basis of our present understanding 
of fundamental matter and interactions.  
In spite of the success of these two theories, there 
is up to now no satisfactory way to reconcile them
in a consistent quantum theory. 
In particular, quantum mechanics combined with
GR strongly suggests a ``foam-like'' or quantum
structure at the Planck scale $10^{-33}$ cm, where
quantum fluctuations of space-time  
are expected to be important.
While some kind of quantum structure of space-time indeed arises
e.g. in string theory, a satisfactory
understanding is still missing.
The cosmological constant problem should 
be seen in the same context:
 the observed tiny (or zero) value of
the cosmological constant 
is in striking contradiction with quantum mechanical expectations, 
which are off by a  factor of order $10^{120}$.
Reconciling quantum mechanics with gravity is therefore
of utmost importance in theoretical physics. 

In view of these problems, it is natural to consider
noncommutative (NC)
or quantum spaces as models for space-time. For fixed backgrounds,
considerable work has been done in this context, leading
to NC field theory  \cite{reviews}. 
Recently, it was understood that gravity emerges naturally from 
NC gauge theory, 
without having to introduce an explicit dynamical metric. 
Earlier forms of this idea \cite{Yang:2006hj,Rivelles:2002ez}
can be cast in concise form for 
matrix models of Yang-Mills type \cite{Steinacker:2007dq},
which describe dynamical quantum spaces. 
We discuss basic results of this approach. 
The IKKT model \cite{Ishibashi:1996xs} 
is singled out as a prime candidate for a 
quantum theory of space-time and matter.

\section{The quantization of Poisson manifolds.}

Space-time in GR is modeled 
by a 4-dimensional manifold $\cM$ with metric $G_{\mu\nu}(x)$.
The basic assumption of the present approach 
is that space-time carries an additional
Poisson structure $\{x^\mu,x^\nu\} =\theta^{\mu\nu}(x)$ (which 
will be related to the metric in \eq{eom-geom-covar-extra}), 
more precisely that space-time is the {\em quantization}
$\cM_\theta$ of such a Poisson manifold. 

In principle, a Poisson structure breaks (local) Lorentz
invariance, which may seem incompatible with observation. 
However, it turns out that $\theta^{\mu\nu}$
does not enter explicitly the effective action of the models
discussed here,
to leading order in an expansion in $\theta^{\mu\nu}$. 
If we assume that the  scale of noncommutativity $\Lambda_{NC}$ defined by
$\det \theta^{\mu\nu} = \Lambda_{NC}^8$ is 
at or near the Planck scale, it is then quite conceivable that its presence
through higher-order terms in the effective action has not been 
detected up to now. 

It is well-known that a Poisson-manifold  
 $(\cM,\theta^{\mu\nu}(x))$ can be quantized 
\cite{Kontsevich:1997vb}. This means that
there exists a quantization map
\be
\begin{array}{rcl}
\cC(\cM) &\to& \cA\,\subset \, L(\cH)\, \\
 f(x) &\mapsto& \hat f(X) 
\end{array}
\label{map} 
\ee
such that $i\{f,g\} \mapsto [\hat f,\hat g] + O(\theta^2)$.
Here $\cC(\cM)$ denotes a suitable space of functions on $\cM$, 
and $\cA$ is interpreted as quantized algebra of 
functions on $\cM$. The quantization of the coordinate functions
$x^\mu$ will be denoted by $X^\mu$. The
matrix model will indeed provide preferred 
coordinates $X^\mu \sim x^\mu$, where
$\sim$ indicates the leading contribution in a semi-classical 
expansion in powers of $\theta^{\mu\nu}$.
These $x^\mu$ are not observable and thus not in conflict with any 
experimental constraints. 
The integral is replaced by the trace, more precisely by the 
volume of the symplectic volume form
\bea
(2\pi)^{2}\, \Tr \hat f &\sim& \int d^{4} x\, \rho(x)\, f \nn\\
\rho(x) &=& (\det\theta^{-1}_{\mu\nu})^{1/2} 
\label{rho-def-general}
\eea
assuming that $\theta^{\mu\nu}$ is non-degenerate.
This is essentially the Bohr-Sommerfeld quantization law.

A simple example of a Poisson manifold  
is given by $\R^4$ together with a constant 
antisymmetric matrix $\{x^\mu,x^\nu\} = \bar\theta^{\mu\nu}$.
Its quantization gives the
Moyal-Weyl quantum plane $\R^4_\theta$, where
\be
[\bar X^\mu,\bar X^\nu] = i \bar\theta^{\mu\nu}\, .
\label{Moyal-Weyl}
\ee
This is formally the (doubled) Heisenberg algebra, i.e.
the usual quantum-mechanical
phase space. However, we need to consider the quantization
of generic Poisson manifold here. 
We only consider the {\em semi-classical} or
geometrical limit of such a quantum space in this paper. This means that 
the space is described in terms of functions on $\cM$ using \eq{map},
keeping only the Poisson bracket on the rhs of \eq{map} and
dropping all higher-order terms in $\theta$. 
Accordingly, we will always replace
$[\hat f(X),\hat g(X)] \to i  \{f(x),g(x)\}$ and 
$[X^\mu,X^\nu] \to i\theta^{\mu\nu}(x)$.
In particular, 
\be
[X^\mu,f(X)] \sim i\theta^{\mu\nu}(x) 
\frac{\partial}{\partial x^\nu} f(x).
\label{derivation}
\ee

\section{Yang-Mills matrix models and their effective geometry}

Consider a matrix model with action
\be
S_{YM} = - \Tr [X^a,X^b] [X^{a'},X^{b'}] 
\eta_{aa'}\eta_{bb'},
\label{YM-action-extra}
\ee
for hermitian matrices 
or operators  $X^a,  \,\, a=1,..., D$ 
acting on some Hilbert space $\cH$.
Here
\be
\eta_{ab'} = {\rm diag}(1,1,...,1) \quad \mbox{or}
 \quad \eta_{ab} = {\rm diag}(-1,1,...,1) 
\label{background-metric}
\ee
in the Euclidean 
resp.  Minkowski case.
The above action is invariant under the following gauge symmetry
\be
X^\mu \to U^{-1} X^\mu U, \qquad U \in U(\cH) .
\label{gauge}
\ee
The equations of motion are
\be
[X^a,[X^b,X^{a'}]] \eta_{aa'} =0 .
\label{matrix-eom}
\ee
The matrix model should be considered as background-independent, since
no  geometry whatsoever is present a priori;
rather, the geometry will arise dynamically.
In particular, 
the model admits as solutions $4$ -dimensional 
noncommutative spaces $\cM_\theta \subset \R^{D}$, 
interpreted as space-time embedded in $D$ dimensions. 
To see this, we split
 the matrices as 
\be
X^a = (X^\mu,\phi^i), \qquad \mu = 1,...,4, 
\,\,\, i=1, ..., D-4
\label{extradim-splitting}
\ee
where the ``scalar fields'' $\phi^i=\phi^i(X^\mu)$ are assumed to 
be functions of $X^\mu$.
The prototype of such a solution is a flat embedding of a
4-dimensional quantum space
\bea
[X^\mu,X^\nu] &=& i\theta^{\mu\nu}, \qquad \mu, \nu= 1,...,4, \nn\\
\phi^i &=& 0,\qquad\quad\,  i=1, ..., D-4
\eea
where $X^\mu$ generates a 4-dimensional NC space $\cM_\theta$;
for example,  $\R^4_\theta\subset \R^{D}$ is realized by 
$[X^\mu,X^\nu] = i\theta^{\mu\nu} \one$.

In general, $\phi^i(X) \sim \phi^i(x)$ will be nontrivial.
One could  interpret $\phi^i(x)$ as scalar fields on 
$\R^4_\theta$; however, it is more appropriate to
interpret $\phi^i(x)$ as purely geometrical degrees of freedom,
defining 
the embedding of a submanifold $\cM \subset
\R^{D}$. 
In the semi-classical limit, suitable ``optimally localized states''
of $\langle X^a\rangle \sim x^a$  
will then be located on $\cM \subset \R^{D}$.
This $\cM$ carries the induced metric 
\be
g_{\mu\nu}(x) = \eta_{\mu\nu} 
+  \partial_{\mu}\phi^i \partial_{\nu}\phi^j\delta_{ij} 
\, = \, \partial_\mu x^a\partial_\nu x^b\, \eta_{ab} 
\label{g-def}
\ee
via pull-back of $\eta_{ab}$. Note that $g_{\mu\nu}(x)$
is {\em not} the metric responsible for gravity, and will enter the
action only implicitly.  Rather, 
it turns out that all fields arising from the matrix model will only
live on the brane $\cM$, and couple to the effective metric 
$G_{\mu\nu}$ given below \eq{G-def-general}.
As opposed to standard braneworld-scenarios,
there really is no higher-dimensional ``bulk''
which could carry physical degrees of freedom here, not even gravitons.

\paragraph{Effective metric and Poisson structure.}

Expressing the $\phi^i$ in terms of $X^\mu$, we obtain
\be
[\phi^i,f(X^\mu)] \,\sim\, i\theta^{\mu\nu}\partial_\mu \phi^i \partial_\nu f
\ee
in the semi-classical limit.
This involves only the components $\mu = 1, ..., 4$ of the
antisymmetric ``tensor'' $[X^a,X^b] \sim i\theta^{ab}(x)$, which has rank $4$
in the semi-classical limit.
Here the derivations
\be
-i[X^\mu,.] \,\sim\, \theta^{\mu\nu} \partial_\nu 
\ee
span the 4-dimensional tangent space of $\cM\subset \R^{D}$, and
define a preferred frame. We can interpret 
\be
[X^\mu,X^\nu] \sim i\theta^{\mu\nu}(x)
\label{theta-induced}
\ee
as Poisson structure on $\cM$, noting that the Jacobi identity
is trivially satisfied. This is  the Poisson structure 
 on $\cM$ whose quantization is given by the matrices 
$X^\mu, \,\, \mu = 1,..., 4$, interpreted as quantization of 
the  coordinate functions $x^\mu$ on $\cM$. 
In particular, the rank of $\theta^{\mu\nu}$ coincides with the 
dimension of $\cM$.
Its inverse $\theta^{-1}_{\mu\nu}(x)$ defines a
symplectic form on $\cM$.

We can now  extract the semi-classical limit
of the matrix model and its physical interpretation.
To understand the effective geometry on $\cM$,
consider a (test-) particle on $\cM$, 
modeled by an additional scalar field $\varphi$.
Due to gauge invariance, the only reasonable
kinetic term is
\bea
S[\varphi] &\equiv& - \Tr [X^a,\varphi][X^b,\varphi] \eta_{ab} = 
- \Tr \left( [X^\mu,\varphi][X^\nu,\varphi] \eta_{\mu\nu} 
  + [\phi^i,\varphi][\phi^j,\varphi] \delta_{ij}\right) 
\label{MM-action-scalar}
\eea
(for example, $\varphi$ could be an $su(n)$ component of $\phi^i$).
In the semi-classical limit, this becomes
\bea
S[\varphi]
&\sim& \frac 1{(2\pi)^2}\, \int d^{4} x\; 
|G_{\mu\nu}|^{1/2}\,G^{\mu\nu}
 \partial_{\mu} \varphi \partial_{\nu} \varphi  
\label{covariant-action-scalar}
\eea
which has the correct covariant form, where\footnote{$G^{\mu\nu}$ is
 corresponds to $\tilde G^{\mu\nu}$ in 
\cite{Steinacker:2008ya,Steinacker:2008ri}} \cite{Steinacker:2008ri}
\be  \fbox{$\quad
G^{\mu\nu}(x) = e^{-\sigma}\, \theta^{\mu\mu'}(x) \theta^{\nu\nu'}(x) 
 g_{\mu'\nu'}(x) 
\label{G-tilde-def}.
\quad  $}
\label{G-def-general}
\ee
Here $g_{\mu\nu}(x)$ 
is the metric \eq{g-def} induced on $\cM\subset \R^{D}$ via 
pull-back 
of $\eta_{ab}$, and
\bea
\qquad \rho &=& (\det \theta^{-1}_{\mu\nu})^{1/2} , 
\qquad e^{-\sigma} = \rho\, |g_{\mu\nu}|^{-\frac 12},  \label{sigma-rho-relation}\\
\eta(x) &=& \frac 14 e^\sigma\, G^{\mu\nu} g_{\mu\nu} .
\label{eta-def}
\eea
Therefore the kinetic term on  $\cM_\theta$
is governed by the metric $G_{\mu\nu}(x)$, 
which depends on the Poisson tensor $\theta^{\mu\nu}$ and the 
embedding metric $g_{\mu\nu}$.
We note that 
\be
|G_{\mu\nu}(x)| = |g_{\mu\nu}(x)|, 
\label{G-g-4D}
\ee
hence $G_{\mu\nu}$ is
unimodular for trivially embedded branes
 (in the preferred matrix coordinates $x^\mu$) \cite{Steinacker:2007dq}. 
Since $\theta^{\mu\nu}$ does 
not enter the Riemannian volume at all, this leads to a
very interesting mechanism for stabilizing flat space, 
which may hold the key for the cosmological constant problem
as discussed below.

\section{Covariant equations of motion.}

We are now in a position to rewrite
the basic matrix e.o.m. \eq{matrix-eom} in a covariant way.
For the scalar fields $\phi^i$, the  e.o.m. is
\bea
0 &=& [X^a, [X^{b},\phi^i]] \eta_{ab} 
= [X^\mu, [X^{\nu},\phi^i]]\eta_{\mu\nu} 
 + [\phi^i, [\phi^j,\phi^i]]\,\delta_{ij} \nn\\
&=&  i[X^\mu, \theta^{\nu\eta}\partial_\eta\phi^i] \eta_{\mu\nu} 
 + i[\phi^i,\theta^{\nu\eta}\partial_\nu\phi^j\partial_\eta\phi^i]\,\delta_{ij} \nn\\
&\sim&  -\theta^{\mu\rho}\partial_\rho
  (\theta^{\nu\eta}\partial_\eta\phi^i) \eta_{\mu\nu} 
 -\theta^{\mu\rho}\partial_\mu \phi^i
 \partial_\rho(\theta^{\nu\eta}\partial_\nu\phi^j\partial_\eta\phi^i) \,\delta_{ij}
\nn\\
&=& e^\sigma (\Gamma^\eta \partial_\eta\phi^i 
  -G^{\rho\eta}\partial_\rho\partial_\eta\phi^i)
= - e^\sigma \Delta_{G} \phi^i \, ;
\label{eom-varphi-0}
\eea
for a detailed derivation see \cite{Steinacker:2008ri}.
The same computation gives 
\be 
\Delta_{G} x^\mu = 0 = \Gamma^\mu,
\label{eom-X-harmonic-tree}
\ee
consistent with the ambiguity of the splitting  
$X^a = (X^\mu,\phi^i)$
into coordinates and scalar fields.
Together with \eq{eom-varphi-0} this implies   
\cite{Steinacker:2008ri} 
\be
\fbox{$
G^{\g \eta}\, \nabla_\g (e^{\sigma} \theta^{-1}_{\eta\nu}) 
\, = \, e^{-\sigma}\,G_{\mu\nu}\,\theta^{\mu\g}\,\partial_\g\eta$}
\label{eom-geom-covar-extra}
\ee
Here $\nabla$ denotes the Levi-Civita connection with respect to
 $G_{\mu\nu}$.
Remarkably, this equation is also a consequence of a ``matrix'' 
Noether theorem due to the  symmetry
$X^a \to X^a + c^a \one$, as shown in  \cite{Steinacker:2008ya}. This
means that equation \eq{eom-geom-covar-extra} is protected from quantum
corrections, and should be taken serious at the quantum level. 
It  has the structure of covariant Maxwell equations,
and gives the relation between the noncommutativity 
$\theta^{\mu\nu}(x)$ and the effective metric $ G^{\mu\nu}$.
For given asymptotical behavior, $\theta^{\mu\nu}(x)$ 
should therefore be completely
determined (apart from gravitational waves, see below) 
by the scalar functions $\eta(x)$ and $e^\sigma$.

We note that \eq{eom-varphi-0} and
\eq{eom-X-harmonic-tree} have a simple interpretation:
the embedding $\cM \subset \R^D$ is harmonic w.r.t.
$G_{\mu\nu}$.
In particular, the dynamical matrices $X^\mu \sim x^\mu$ define
harmonic coordinates for on-shell geometries, which
in general relativity 
would be interpreted as gauge condition. However,
\eq{eom-varphi-0} and \eq{eom-X-harmonic-tree} individually
might be subject to quantum corrections
as indicated below.

The main message is that the kinetic term in the matrix model 
necessarily 
involves the metric $G_{\mu\nu}(x)$.
This also holds for gauge fields and fermions,
possibly up to density factors.
Therefore $G_{\mu\nu}$ must be interpreted as gravitational
metric. 
It is dynamical, because it depends on 
the embedding fields $\phi^i$ and the Poisson structure
$\theta^{\mu\nu}$,
both of which are dynamical. Using a standard embedding theorem 
\cite{clarke}, it follows that $G_{\mu\nu}$ can describe in
principle the most general metric in 4 dimensions for the 
case $D\geq 10$ (ignoring the e.o.m. for now).
Therefore the matrix model provides a non-perturbative
definition of a 
theory of space-time and gravity, which 
naturally includes also gauge fields and matter.
The quantization can also be performed
relatively easily in a perturbative manner, taking advantage 
of an alternative interpretation as noncommutative gauge theory:
the dynamical Poisson structure can be parametrized in terms of a 
$u(1)$ gauge field as indicated below. This leads to the 
hope that this type of matrix model may serve as a 
model for quantum gravity, quite possibly more accessible than
other, more conventional approaches.

\section{Dynamical emergent gravity}

Before making contact with real physics, we need to understand
the dynamics of emergent gravity. Notice that 
we have not encountered anything like the Einstein-Hilbert
action, and in fact it seems impossible to simply 
add such a term in the matrix model.
This is of course extremely interesting: either the model
will fail hopelessly, or it will provide a very remarkable
and predictive new mechanism for gravity, free of theoretical 
prejudice.

We can give at least 2 arguments which  suggest that 
the dynamics of the geometry
resp. gravity in the matrix model should be quite close to GR: 

\begin{enumerate}
\item
the dynamics of $\theta^{\mu\nu}(x)$ 
as given by the matrix model indeed implies $R_{\mu\nu} =0$
at least for linearized metric fluctuations \eq{h-abdown} around 
flat Moyal-Weyl space without matter. 
Moreover, it provides the 2 physical degrees of 
freedom for gravitons \eq{h-abdown}. This is a remarkable observation, 
essentially due to Rivelles \cite{Rivelles:2002ez}; 
see \cite{Steinacker:2008ya} for more details.
\item
as soon as the matrix model
is quantized, the effective action will
 contain an induced Einstein-Hilbert action.
This is discussed below in more detail.
\end{enumerate}

\subsection{Quantization and induced gravity}

The quantization of the matrix model \eq{YM-action-extra} is defined by 
\be
Z = \int d X^a e^{-S_{YM}[X]}\, ,
\ee
which is easily generalized to include fermions,
notably in the IKKT model \cite{Ishibashi:1996xs} for $D=10$.
As shown in \cite{Austing:2001pk}, this integral is well-defined for
finite $N$ (in $D\geq 3$ dimensions), apart from the flat directions 
corresponding to $X^a \to X^a + c^a \one$ which could 
be handled using standard methods.
This provides a non-perturbative definition of the model
at the quantum level. To gain some insight, we consider a 
perturbative treatment around a given background as discussed
above.
For simplicity, consider the quantization of an additional scalar field
$\varphi$ coupled to the matrix model as in \eq{MM-action-scalar}, which 
upon integration leads to an effective action 
\be
e^{-\Gamma_{\varphi}} = \int d\varphi\, e^{-S[\varphi]} , \quad\mbox{where}\quad
 \Gamma_{\varphi} = \frac 12 \Tr \log \Delta_{G} \, .
\ee
A standard argument using the heat kernel expansion
of $\Delta_{G}$ gives
\be
\Gamma_{\varphi} = \frac 1{16\pi^2}\, \int d^4 x 
\sqrt{|G_{\mu\nu}|}\,\left( c_1\Lambda_1^4 
+ c_4 R[G]\, \Lambda_4^2 + O(\log \Lambda) \right) \, ,
\label{S-oneloop-scalar}
\ee
which is also the general structure of the 1-loop effective action 
for the geometrical sector due to other fields. 
The coefficients $c_i$ as well as the effective cutoffs $\Lambda_i$
depend on the detailed
field content of the model, and can be obtained essentially from
Seeley-de Witt coefficients\footnote{however the contributions from
  fermions and the ``would-be $U(1)$ fields'' are non-standard as
  discussed in \cite{Klammer:2008df}.}.
This is essentially the mechanism of induced gravity 
\cite{Sakharov:1967pk}.

The action \eq{S-oneloop-scalar} suggests to relate the cutoff
$\Lambda_4^2$ with the gravitational constant $\frac 1G$.
On the other hand, $\Lambda_4$ is expected to be the scale 
of $N=4$ SUSY breaking, since only $N=4$ supersymmetric 
models do not induce this term \cite{Matusis:2000jf,Klammer:2008df}; 
this is related to the 
lack of UV/IR mixing in these models as discussed below.
This suggests that 
in order to have a finite gravitational coupling constant, the 
model should  have $N=4$ supersymmetry above 
a certain scale. This is realized in the IKKT model. 
The scale $\Lambda_1$ may be different from $\Lambda_4$.

Note that a term $\int d^4 x\sqrt{G}\, \Lambda^4$
is usually interpreted as cosmological constant, and its 
scaling with $\Lambda^4$ usually presents a major problem. 
This is nothing but the 
cosmological constant problem, which represents arguably the biggest
challenge to our understanding of gravity.
We claim that this problem may be resolved here.
To see this, recall 
$|G_{\mu\nu}| =|g_{\mu\nu}|$  \eq{G-g-4D}, 
independent of $\theta^{\mu\nu}(x)$.
Moreover, 
$$
\d \int d^4 x\sqrt{G} \, \sim \,
\int d^4 x\sqrt{g} g^{\mu\nu}\d g_{\mu\nu} \, \sim \, 
\int d^4 x\sqrt{g} \d \phi^i \Delta_g \phi^j \delta_{ij} 
$$ 
vanishes for 
harmonic embeddings $\Delta_g \phi^i=0$. 
Therefore for such backgrounds, this ``would-be cosmological constant
term'' is irrelevant and  does not enter the equations of motion
for the geometry.
For example, flat (Moyal-Weyl) space is a solution even 
at one loop\footnote{and presumably to all loops;
this can also be seen from the point
of view of $U(1)$ gauge theory discussed below.},
{\em without fine-tuning} $\Lambda$.
Therefore minimally or harmonically embedded branes 
(w.r.t. $g_{\mu\nu}$)
are {\em protected from the  cosmological constant problem} 
\cite{Steinacker:2008ri};
the term $\int d^4 x\sqrt{G}\, \Lambda^4$ is present but simply 
does not imply any cosmological constant. This is due to the 
particular parametrization of the geometry in terms of 
$\theta^{\mu\nu}$ and $\phi^i$ rather than a
fundamental metric.
Notice  that $g_{\mu\nu}$ does not necessarily (but quite possibly) 
coincide with the effective metric $G_{\mu\nu}$. This makes the 
full analysis of the gravitational sector in the matrix model
quite non-trivial, and much more work is required.
In view of the recently found  cosmological solutions 
 \cite{cosm-solution}, the present model may provide a serious 
candidate for resolving the cosmological constant problem 
without fine-tuning, going beyond GR.

\section{Gauge fields}

We now relate the above discussion
with noncommutative gauge theory. The main result is that 
the matrix model naturally includes 
$su(n)$ gauge theory coupled to gravity. If the 
same model is viewed as $u(n)$ gauge theory on $\R^4_\theta$, then the 
trace-$u(1)$ sector is afflicted by the notorious 
UV/IR mixing. This is nothing but a reflection of the 
induced gravitational action.

\subsection{Would-be $u(1)$ gauge fields}

Recall that
a particular solution of the matrix equations of motion 
\eq{matrix-eom} is given by the generators $\bar X^{\mu}$ of the 
Moyal-Weyl quantum plane $\R^4_\theta$. Its
effective geometry is indeed flat, given by
\be
\bar G^{\mu\nu} = \bar\rho\,\bar\theta^{\mu\mu'}\,\bar\theta^{\nu\nu'}
      g_{\mu'\nu'}\, 
\qquad \bar\rho = (\det\bar\theta^{\mu\nu})^{-1/2} \equiv \Lambda_{NC}^4
\label{effective-metric-bar}
\ee
assuming trivial embedding.
Fluctuations of $X^a$ can be parametrized in terms of 
$u(1)$ gauge fields $A_\mu$ and scalar fields $\phi^i$ on 
 $\R^4_\theta$, using 
\bea
X^a &=& (\bar X^{\mu} + \cA^\mu\, , \phi^i) \nn\nn\\
\,[\bar X^\mu + \cA^\mu,f] &=& i \bar\theta^{\mu\nu} 
(\frac{\partial}{\partial\bar x^\nu} 
+ i [A_\nu,.]) f \, \equiv \, i \bar\theta^{\mu\nu} D_\nu f 
\eea
where $\cA^a \sim \cA^a(\bar x) \equiv -\bar\theta^{ab} A_b(\bar x)$.
The actions \eq{YM-action-extra} could then be written as 
$$
S_{YM} = \int d^4 \bar x\,\left(\bar \rho^{-1}\, \bar G^{\mu\mu'}\,\bar
G^{\nu\nu'}\,F_{\mu\nu}\,F_{\mu'\nu'}
+ 2 \bar G^{\mu\nu}\, D_\mu\phi^i D_\nu \phi^i \d_{ij}
- [\phi^i,\phi^j][\phi^{i'},\phi^{j'}] \d_{ij} \d_{i'j'}
\, + \bar G^{\mu\nu} g_{\mu\nu}\right)  
$$
where $F_{\mu\nu} = \partial_\mu A_\nu - \partial_\nu A_\mu  + i[A_\mu,A_\nu]\,$ 
is the
$u(1)$ field strength. This might suggest an interpretation as
noncommutative $u(1)$ gauge theory on $\R^4_{\bar \theta}$. 
However, $A_\mu(\bar x)$ is completely
absorbed in the metric $G^{\mu\nu}(x)$ resp. $\theta^{\mu\nu}(x)$
in the proper geometrical interpretation, via
\be
\theta^{\mu\nu}(x) = \bar\theta^{\mu\nu}(\bar x)
-  \bar\theta^{\mu\mu'}\bar\theta^{\nu\nu'} F_{\mu'\nu'} .
\label{theta-F}
\ee
This 
explains why these would-be $u(1)$ gauge fields cannot be disentangled
e.g. from the $su(n)$ components: all fields
couple to the metric $G_{\mu\nu}$, which is 
a function of $F_{\mu\nu}$ via \eq{theta-F}. 
In particular, an
induced gravitational action \eq{S-oneloop-scalar} arises 
upon quantization, which does 
{\em not} simply renormalize 
the tree-level action. This is precisely the ``strange'' UV/IR 
contribution \cite{Minwalla:1999px}
 to the effective action for the would-be $u(1)$ gauge fields
 in the IR limit \cite{Grosse:2008xr}.

Expanding the effective metric 
$G_{\mu\nu} = \bar G_{\mu\nu} + h_{\mu\nu}$ 
around the flat Moyal case to leading
order in $F_{\mu\nu}$ leads to an expression for the linearized
metric fluctuations
\be
h_{\mu\nu}  =  - \bar G_{\nu \nu'} \bar\theta^{\nu'\rho} \bar F_{\rho\mu} 
- \bar G_{\mu \mu'} \bar\theta^{\mu'\rho} \bar F_{\rho\nu} \, 
- \frac 12 \bar G_{\mu \nu} \bar F_{\rho\eta} \bar\theta^{\rho\eta}
\label{h-abdown}
\ee
Remarkably, the equations of motion of the matrix model
imply that the vacuum geometries are Ricci-flat \cite{Rivelles:2002ez}, 
\be
R_{\mu\nu}[\bar G + h] \, =\, 0 \quad + O(\theta^2),
\qquad \mbox{while}\quad R_{\mu\nu\rho\sigma} \neq 0\, .
\label{R-flat}
\ee

\subsection{Nonabelian gauge fields}
\label{sec:nonabelian}

Finally we show how $su(n)$-valued gauge fields arise 
in the same matrix model, on
a suitable background. To avoid confusion we denote such a 
``nonabelian'' background with
\be
Y^a = \left(\begin{array}{l}Y^\mu \\ Y^i \end{array}\right) = 
\left\{\begin{array}{ll} X^\mu\otimes \one_n,  & \quad a=\mu = 1,..., 4,   \\ 
\phi^i \otimes \one_n, & 
\quad a = 4+i, \,\, i = 1, ..., D-4 . \end{array} \right.
\ee
This can be understood as $n$ copies of the brane configurations
considered above. 
We want to understand general fluctuations around this new background.
Since the $u(1)$ components describe the geometry, 
we expect to find $su(n)$-valued gauge fields as well as
 scalar fields in the adjoint.
It turns out that the following gives an appropriate parametrization 
of these general fluctuations:
\be
\left(\begin{array}{l}Y^\mu \\ Y^i \end{array}\right) = 
\left(\begin{array}{l} X^\mu\otimes \one_n + \cA^\mu   \\ 
\phi^i\otimes \one_n  + \Phi^i
+ \cA^\rho \partial_\rho (\phi^i\otimes \one_n  + \Phi^i)\end{array} \right)
\label{nonabelian-SW}
\ee
where
\bea
\cA^\mu &=&  \cA^\mu_\a \otimes \lambda^\a
\,\, = \,\, - \theta^{\mu\nu} A_{\nu,\a} \otimes \lambda^\a, \nn\\
\Phi^i  &=&  \Phi^i_\a\otimes \lambda^\a 
\eea
parametrize the $su(n)$-valued gauge fields resp. scalar fields, and
$\lambda^\a$ denotes the generators of $su(n)$.
This amounts to the 
leading term in a Seiberg-Witten map 
\cite{Seiberg:1999vs}, relating 
noncommutative and commutative $su(n)$ gauge fields.

One can now show that the semi-classical 
limit of the matrix model action \eq{YM-action-extra}
for these $su(n)$-valued 
gauge fields $A_{\mu}$ on general 4-dimensional $\cM_\theta\subset\R^D$
is given by
\bea
S_{YM}[\cA]  
&\sim&  \int d^{4} x\, |G_{\mu\nu}|^{1/2}
e^{\sigma} \,  G^{\mu\mu'}  G^{\nu\nu'}
\tr(F_{\mu\nu}\,F_{\mu'\nu'})\,\, 
+ 2  \int  \eta(x) \, \tr F\wedge F  \; .
\label{S-NC-def}
\eea
where $F_{\mu\nu} = \partial_\mu A_\nu - \partial_\nu A_\mu + i
[A_\mu,A_\nu]$
is the $su(n)$-valued field strength.
This was shown in  \cite{Steinacker:2007dq} for $D=4$ using a
direct but long computation, and 
in \cite{Steinacker:2008ya} for the general case via the 
equations of motion. 
A similar result applies  for higher-dimensional branes.
Remarkably, the corresponding Yang-Mills equations of motion
are a direct consequence of a matrix Noether theorem
corresponding to the symmetry $X^a \to X^a + c^a \one$ of the matrix
model \cite{Steinacker:2008ya}.

Finally, fermions are naturally included in these models 
\cite{Ishibashi:1996xs,Steinacker:2008ri} and turn out to couple to the 
same metric $G_{\mu\nu}$, 
albeit possibly (depending on the geometry) with a 
non-standard spin connection; 
see \cite{Klammer:2008df} for results on the $D=4$ case.

We conclude that
Yang-Mills matrix models are strong candidates
for a unified theory of fundamental interactions, and
promise advantages over GR for quantization and the 
cosmological constant problem.
A simple and intrinsically noncommutative 
mechanism for gravity is identified.
However, more work
is required to obtain a thorough understanding and judgment.

\vspace{0.3cm}
{\bf Acknowledgments} This work was supported by 
FWF project P20017.

\end{document}